\documentclass[showpacs,twocolumn]{revtex4}
\usepackage{bm}
\usepackage{graphicx}
\usepackage{amsmath}
\usepackage{amssymb}

\begin{document}
\title{Spin-orbit interaction and weak localization in heterostructures}
\author{M.\,M.~Glazov} 
\author{L.\,E.~Golub}
\address{Ioffe Physical-Technical Institute of the Russian
Academy of Sciences, 194021 St.~Petersburg, Russia}

\begin{abstract}

Theory of weak localization in two-dimensional high-mobility semiconductor systems is developed with allowance for the spin-orbit interaction. The obtained expressions for anomalous magnetoresistance are valid in the whole range of classically weak magnetic fields and for arbitrary strengths of bulk and structural inversion asymmetry contributions to the spin splitting. The  theory serves for both diffusive and ballistic regimes of electron propagation taking into account  coherent backscattering and nonbackscattering processes. The transition between weak localization and antilocalization regimes is analyzed. The manifestation of the mutual compensation of Rashba and Dresselhaus spin splittings in magnetoresistance is discussed. Perfect description of experimental data on anomalous magnetoresistance in high-mobility heterostructures is demonstrated. The in-plane magnetic field dependence of the conductivity caused by an interplay of the spin-orbit splittings and Zeeman effect is described theoretically.
\end{abstract}


\maketitle

\section{Introduction}

The electron, being a quantum object, manifests both particle and wave properties while propagating in a solid. In the state-of-the-art semiconductor heterostructures where the electron mean free path $l$ caused by scattering from remote impurities, phonons and interface imperfections exceeds by far the electron wavelength $2\pi/k_{\rm F}$ ($k_{\rm F}$ is Fermi wavevector) the electron transport is known to be primarily classical and described by Drude theory. However, a quantum nature of an electron is clearly demonstrated by a variation of the conductivity at low temperatures with a small magnetic field  perpendicular to the system plane.
Low-field magnetoresistance is caused by the \textit{weak localization} of electron waves: a particle
can propagate via different paths, and among them there is a number of self-crossing paths with loops. An electron can pass a loop by two trajectories: clockwise and counter-clockwise, which leads to a constructive interference resulting in the increase of the probability for the particle to return to the initial point. This increase of the return probability means an increase of resistance and decrease of the conductivity in comparison to the Drude formula. Classically small magnetic field causes an electron to obtain a phase difference equal to the magnetic flux through the loop for propagation clockwise and counterclockwise. Thus application of a perpendicular magnetic field suppresses the constructive interference and increases the conductivity. Due to unusual field dependence this phenomenon is known as \textit{anomalous negative magnetoresistance} or positive magnetoconductivity~\cite{AA}.

Crucial effect on the above picture makes a spin-orbit interaction.
If it is strong, the electronic waves also interfere after passage the loops in two opposite directions, but this interference is destructive. As a result, the probability of return is smaller than the classical value, so the conductivity correction is positive. Perpendicular magnetic field destroys this interference as well but in this case it leads to {a} decrease of conductivity, i.e. to \textit{positive magnetoresistance}. Since the {situation} is totally opposite to the spinless case the interference effect in the presence of spin-orbit interaction is called \textit{weak antilocalization}. 

In semiconductor heterostructures the spin-orbit interaction
is described by the following Hamiltonian~\cite{SST_BPZ}
\begin{equation}\label{H_SO}
    H(\bm{k}) = \hbar  \: \bm{\sigma} \cdot
    \bm{\Omega}(\bm{k}),
\end{equation}
where $\bm{k}$ is the electron wave vector, $\bm{\sigma}$ is the
vector of Pauli matrices, and $\bm{\Omega}(\bm k)$ is an odd function of
$\bm k$. The spin splitting due to the spin-orbit interaction
Eq.~(\ref{H_SO}) equals to $2 \hbar \Omega({\bm k})$.

Theory of weak-antilocalization induced alternating magnetoresistance has been developed by \textit{Pikus~et~al.} in the middle of 1990s~\cite{ILP,PikusPikus}. 
It had successfully described weak-antilocalization experiments on available in that time low-mobility heterostructures~\cite{Knap}. 
However the obtained expressions are
valid only for weak spin-orbit interaction and very low
magnetic fields. The former assumption means that $\Omega \tau \ll
1$, where $\tau$ is the scattering time, and the latter condition
reads as $l_B \gg l$, where $l_B = \sqrt{\hbar / e B}$ is the
magnetic length. This so-called
``diffusive'' regime takes place in fields $B \ll B_{tr}$, where
$$B_{tr} = {\hbar \over 2 e l^2}$$ is the ``transport'' field at which the magnetic length $l_B$ equals to the mean free path $l$. In the theory~\cite{ILP,PikusPikus} $B_{tr}$ is assumed to be infinitely large which is a good approximation for low-mobility samples.

However, starting from the {early} 2000s, anomalous magnetoresistance measurements are being performed on  high-mobility samples in different laboratories around the world, see e.g.~\cite{exp_do1,exp_do2,exp_do3,exp_do4}. 
The motion of the particle on the trajectories relevant for the interference in such systems is ballistic rather than diffusive. The field $B_{tr}$ is small in these structures being less than 1~mT. The characteristic magnetoresistance maximum occurs at $B>B_{tr}$, i.e. out of the range of applicability of the theory existed that time. Moreover the systems started to appear having large spin-orbit splitting~\eqref{H_SO} and long scattering times so the product $\Omega \tau$ is even larger than unity. The question to theory was sharply raised after publication of the paper by \textit{Studenikin~et~al.}~\cite{exp_do2}: it has been demonstrated that both low-field and high-field parts of the magnetoresistance curve can be fitted by the theory~\cite{ILP} but with absolutely different sets of fitting parameters.   After the paper~\cite{exp_do2} it became especially clear that a new theory of weak localization is required.

Such theory has been recently developed in Refs.~\cite{PRB05,FTP06}.
The obtained expressions for the magnetoconductivity are valid in the whole range of classically weak magnetic fields and for any values of the spin splittings, i.e. for
arbitrary values of $B/B_{tr}$ and $\Omega \tau$. The  theory takes into account low symmetry of $[001]$ grown heterostructures where both bulk and structure inversion asymmetry contributions to $\bm \Omega(\bm k)$ in Eq.~\eqref{H_SO} with linear and cubic in the wavevector terms {coexist}. This theory opened a possibility to describe anomalous magnetoresistance experiments
and to extract adequately spin-splitting and kinetic parameters of high-mobility two-dimensional (2D) semiconductor systems.

In the magnetic field normal to the heterostructure plane only orbital effects  are important in the
anomalous magnetoresistance while the Zeeman splitting plays no role. 
Another interesting possibility opens up in the case of the magnetic field applied in the plane of the heterostructure. In such a case the orbital effect of the magnetic field is relatively unimportant while the Zeeman splitting dramatically affects weak localization. 
In strong in-plane magnetic fields the Zeeman effect completely overcomes spin-orbit effect and restores the spin-orbit-less value of the quantum conductivity correction. However in the intermediate regime the magnetoresistance is formed as a result of interplay of the Zeeman and spin-orbit splittings. The in-plane magnetoresistance is not sufficiently studied at present.

In this review we describe the weak-antilocalization theory for high-mobility 2D semiconductor systems and present the expressions for anomalous magnetoconductivity valid in the whole range of classically-weak fields and for arbitrary large spin-orbit splitting. 
We demonstrate that this theory perfectly describes the experimental data on anomalous magnetoresistance of high-mobility 
heterostructures.
We also put forward a theory describing quantum corrections to the conductivity due to  an interplay of spin-orbit effects and Zeeman effect of arbitrary strengths caused by an in-plane magnetic field.

\section{Theory}

There are two contributions of different nature to the spin-orbit
interaction Hamiltonian Eq.~(\ref{H_SO}) in 2D semiconductor
systems: the Rashba term $\bm{\Omega}_R$ and the Dresselhaus term
$\bm{\Omega}_D$. In heterostructures grown along the direction $z
\parallel [001]$ both vectors $\bm{\Omega}_{R}$ and $\bm{\Omega}_D$ lie in the 2D plane. {The Rashba term contains only first angular harmonics of the wavevector, while the Dresselhaus term contains both first and third harmonics. They} have
the following form
\begin{equation}\label{Omega}
\bm \Omega(\bm k) = \bm \Omega^{(1)}(\bm k) + \bm \Omega^{(3)}(\bm k),
\end{equation}
\[
\bm \Omega_R^{(1)}(\bm k) = \Omega_R (\sin{\varphi}, - \cos{\varphi}),\nonumber
\]
\[
\bm \Omega^{(1)}_D(\bm k) = \Omega_D (\cos{\varphi}, - \sin{\varphi}), \nonumber
\]
\[
\bm \Omega_D^{(3)}(\bm k) = \Omega_{D3} (\cos{3\varphi}, \sin{3\varphi}),
\]
where $\varphi$ is an angle between $\bm k$ and the axis $x\parallel [100]$.

We consider low magnetic fields
\[    
\omega_c \ll \Omega, \: \tau^{-1} \ll E_{\rm F}/\hbar,
\]
where $\omega_c$ is the cyclotron frequency, and $E_{\rm F}$ is the Fermi energy, i.e. those fields where the cyclotron motion of electrons in unimportant. The conductivity correction due to weak localization is given
by a sum of two terms
$$\sigma(B) = \sigma_a + \sigma_b,$$
where $\sigma_a$ and $\sigma_b$ can be interpreted as
backscattering and nonbackscattering interference corrections to
conductivity. In the case of isotropic scattering they are given by~\cite{PRB05,FTP06}
\begin{equation}\label{sigma_a}
    \sigma_a = - {e^2 \over 2 \pi^2 \hbar} \left( {l \over l_B}
    \right)^2 \Biggl\{ {\rm Tr} \left[ \mathcal A^3 ({\cal I} - \mathcal A)^{-1} \right]
    - \sum_{N=0}^\infty {P_N^3 \over 1 - P_N}\Biggr\},
\end{equation}

\begin{eqnarray}\label{sigma_b}
    \sigma_b = {e^2 \over \pi^2 \hbar} \left( {l \over l_B}
    \right)^2
    \Biggl\{ {\rm Tr}
    \left[ \mathcal K^2\mathcal A ({\cal I} - \mathcal A)^{-1}
    \right]\\
       - \frac{1}{4}\sum_{N=0}^\infty Q_N^2\left({P_N \over 1 - P_N} + {P_{N+1} \over 1 - P_{N+1}}\right)
   \Biggr\},\nonumber
\end{eqnarray}
where
$$    P_N = {l_B \over l} \int\limits_0^\infty dx
    \exp{\left( -x {l_B \over \tilde{l}} - {x^2 \over 2}\right)}
    L_N(x^2),
$$
\begin{eqnarray}
    &Q_N&= {1 \over \sqrt{N+1}} {l_B \over l} \nonumber\\
    &\times& \int\limits_0^\infty dx
    \exp{\left( -x {l_B \over l} - {x^2 \over 2}\right)}
    x L_N^1(x^2),\nonumber
\end{eqnarray}
with $\tilde{l} = l/ (1 + \tau/\tau_\phi)$, $\tau_\phi$ being the dephasing time,
$L_N^m$ are the associated Laguerre polynomials, and ${\cal I}$ is the unit operator.
The matrix elements of the operators $\mathcal A$ and $\mathcal K$ in the basis of Landau levels of a particle with a charge $2e$ and in the representation of the total momentum of interfering particles $\bm{S}$ ($S=1$, the momentum projection $m=1,0,-1$) are given by
\begin{eqnarray}\label{A_NN1}
\mathcal A(N,m;N',m') &=&  \int d^2{R} \: {\exp{(-R/\tilde{l})} \over 2\pi R l} F_{NN'}(\bm{R}) \\
&\times& \langle m'|\exp{[ - 2{\rm i} \tau \bm{S} \cdot
\bm{\omega}(\bm{R})]}|m \rangle, \nonumber
\end{eqnarray}
\[
F_{NN'}(\bm{R}) = e^{-t^2 / 2} \: L_N^{N'-N}(t^2) \: (-t e^{{\rm
i} \vartheta})^{N'-N} \: \sqrt{N! \over N'!}.
\]
Here $\bm \omega(\bm R) = \bm \Omega(k_{\rm F}\bm n) R/l$, where $\bm n =(\cos\vartheta,\sin\vartheta)$ is a unit vector pointing along $\bm R$,
% $\vartheta$ is an angle between $\bm R$ and the $x$ axis, 
and $t=R/l_B$. The expression for matrix elements $\mathcal K(N,m;N',m')$ are different from Eq.~\eqref{A_NN1} by the additional factor ${\rm i}\cos{\vartheta}$ in the integrand. In Eqs.~\eqref{sigma_a} and~\eqref{sigma_b} the trace, $\rm Tr$, is a sum of matrix elements with $m=m'$. 

\section{Results and Discussion}

In this section we consider consequently the weak antilocalization caused by isotropic spin-orbit splitting, interplay of Rashba and linear Dresselhaus splittings, cubic in $\bm k$ Dresselhaus splitting, and present comparison of the theory with experimental data.

\subsection{Isotropic spin splitting}

First, we discuss the Rashba or linear Dresselhaus spin-orbit interaction dominance. In both cases the spin  splitting $2\hbar\Omega({\bm k})$ is isotropic in ${\bm k}$-space. This makes possible a partial diagonalization of the operators $\mathcal A$ and $\mathcal K$. For Rashba spin-orbit interaction, it takes place in the basis of
the states $|N, m\rangle$ with equal $N+m$: $|N-2, 1\rangle$,
$|N-1, 0\rangle$, $|N, -1\rangle$, while for Dresselhaus term this
takes place for the states with equal $N-m$. In both cases
we have~\cite{FTP06}:
\begin{equation}\label{sigma_a_fin}
    \sigma_a = - {e^2 \over 2 \pi^2 \hbar} \left( {l \over l_B}
    \right)^2\sum_{N=0}^\infty \Biggl\{ {\rm Tr} \left[ A_N^3 (I - A_N)^{-1} \right]
    - {P_N^3 \over 1 - P_N}\Biggr\},
\end{equation}

\begin{eqnarray}\label{sigma_b_fin}
    \sigma_b = {e^2 \over 4 \pi^2 \hbar} \left( {l \over l_B}
    \right)^2 \sum_{N=0}^\infty
    \Biggl\{ {\rm Tr}
    \left[ K_N {K}_N^T A_N (I - A_N)^{-1} \right] \\
     + {\rm Tr}
     \left[ {K}_N^T K_N A_{N+1} (I - A_{N+1})^{-1} \right] \nonumber \\
   - Q_N^2\left({P_N \over 1 - P_N} + {P_{N+1} \over 1 - P_{N+1}}\right)
   \Biggr\}. \nonumber
\end{eqnarray}

\begin{figure}
\includegraphics[width=\linewidth]{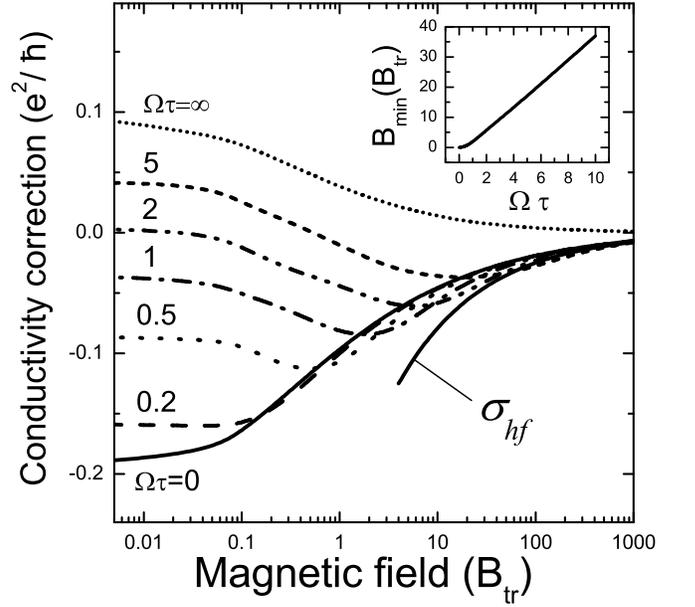}
\caption{Quantum conductivity correction for different strengths of
spin-orbit interaction at $\tau/\tau_\phi = 0.01$. The inset
represents the positions of minima in the magnetoconductivity. After~\cite{PRB05}.}
 \label{IsotrSS}
\end{figure}

Here $I$ is $3\times 3$ unit matrix,
\begin{equation}\label{A_N}
    A_N = \left(%
\begin{array}{ccc}
  P_{N-2} - S_{N-2}^{(0)} & R_{N-2}^{(1)} & S_{N-2}^{(2)} \\ \nonumber
  R_{N-2}^{(1)} & P_{N-1} - 2S_{N-1}^{(0)} & R_{N-1}^{(1)} \\ \nonumber
  S_{N-2}^{(2)} & R_{N-1}^{(1)} & P_N - S_N^{(0)} \\
\end{array}%
\right),
\end{equation}
\begin{equation}\label{K_N}
    K_N = \left(%
\begin{array}{ccc}
  Q_{N-2} - S_{N-2}^{(1)} & R_{N-2}^{(2)} & S_{N-2}^{(3)} \\ \nonumber
  -R_{N-1}^{(0)} & Q_{N-1} - 2S_{N-1}^{(1)} & R_{N-1}^{(2)} \\ \nonumber
  -S_{N-1}^{(1)} & -R_{N}^{(0)} & Q_N - S_N^{(1)} \\
\end{array}%
\right),
\end{equation}

\begin{eqnarray}\label{S_N}
    &&S_N^{(m)}= {l_B \over l} \sqrt{N! \over (N+m)!}    \nonumber\\
    &\times& \int\limits_0^\infty dx
    \exp{\left( -x {l_B \over l} - {x^2 \over 2}\right)}
    x^m L_N^m(x^2) \sin^2\left(\Omega \tau {l_B \over l} x \right),
    \nonumber
\end{eqnarray}
\begin{eqnarray}\label{R_N}
    &&R_N^{(m)}= {l_B \over l \sqrt{2}} \sqrt{N! \over (N+m)!} \nonumber\\
    &\times& \int\limits_0^\infty dx
    \exp{\left( -x {l_B \over l} - {x^2 \over 2}\right)}
    x^m L_N^m(x^2) \sin\left(2\Omega \tau {l_B \over l} x \right).
    \nonumber
\end{eqnarray}
Note that the values with negative indices appearing  in above 
equations for $A_N$ and $K_N$ at $N=0,1$
should be replaced by zeros.

Equations~(\ref{sigma_a_fin}) and~(\ref{sigma_b_fin}) yield the
weak-antilocalization correction to the conductivity in the whole
range of classically-weak magnetic fields and for arbitrary values
of $\Omega \tau$.
In the limit of zero spin splitting, 
\begin{equation}\label{sigma_a_no_spin}
    \sigma_a = - {e^2 \over \pi^2 \hbar} \left( {l \over l_B}
    \right)^2\sum_{N=0}^\infty {P_N^3 \over 1 - P_N},
\end{equation}
\begin{equation}\label{sigma_b_no_spin}
    \sigma_b = {e^2 \over 2 \pi^2 \hbar} \left( {l \over l_B}
    \right)^2\sum_{N=0}^\infty
    Q_N^2 \left({P_N \over 1 - P_N} + {P_{N+1} \over 1 - P_{N+1}}\right).
\end{equation}
Equations~(\ref{sigma_a_no_spin}) and~(\ref{sigma_b_no_spin}) were obtained as results of non-diffusive theory developed for $\Omega = 0$ in Ref.~\cite{Zyuzin}.
In a magnetic field $B \gg (\Omega \tau)^2 B_{tr}$, the
conductivity is independent of $\Omega$, and it is also described by the Eqs.~\eqref{sigma_a_no_spin}, \eqref{sigma_b_no_spin}. The reason is that
in so strong field the dephasing length due to magnetic field
$\sim l_B$ is smaller than one due to spin-orbit interaction, $l /
\Omega \tau$. As a result, the particle spins keep safe at
characteristic trajectories. The conductivity for any finite
$\Omega\tau$ has the zero-$\Omega$ asymptotic. For
$\Omega\tau < 1$ this dependence is achieved at $B \lesssim
B_{tr}$. In high magnetic field $B \gg B_{tr}, (\Omega \tau)^2
B_{tr}$, the conductivity correction has the high-field
asymptotic~\cite{Zyuzin}
\begin{equation}
\label{hf_asympt}
	\sigma_{hf}(B) = - 0.25\sqrt{B_{tr} \over B} \: {e^2 \over \hbar}.
\end{equation}

In Fig.~\ref{IsotrSS} the conductivity correction is plotted for
different strengths of spin-orbit interaction. 
The non-monotonous dependence $\sigma(B)$ can be qualitatively explained by noting that the spin state of two interfering electrons can be either triplet or singlet~\cite{AA}. Indeed, the singlet configuration corresponding to the total spin $S=0$ is unaffected by the spin-orbit interaction while the triplet contribution is suppressed. These spin states contribute to the conductivity correction with opposite signs: singlet contribution is positive while triplet one is negative. The small magnetic field suppresses singlet term only, leading to the decrease of the conductivity while in higher fields both singlet and triplet states are suppressed. Therefore, the magnetoconductivity in high fields is positive. As a result the conductivity as a function of the magnetic field is non-monotonic with a minimum at a certain value of the field, $B_{min}$.

Figure~\ref{IsotrSS} shows that
for $\Omega\tau \lesssim 1$, $\sigma(B)$ coincides with the zero-$\Omega$
dependence for $B > B_{min}$. The asymptotic $\sigma_{hf}(B)$ is
reached at $B \approx 100~B_{tr}$ for all {presented} values of $\Omega
\tau$. The positions of minima in the curves are shown in the
inset. One can see that $B_{min}$ almost linearly depends on the
spin splitting at $\Omega \tau > 0.8$. Fitting yields the
following approximate law
$$B_{min} \approx (3.9 \: \Omega \tau -2) B_{tr}.$$

In the limit $\Omega\tau \rightarrow \infty$
%, the triplet state with $m_s = 0$ does not contribute to the conductivity. The corresponding dependence is presented in Fig.~\ref{IsotrSS}. 
one can see a decrease of conductivity in the whole range of magnetic
fields. At $B \gg B_{tr}$, the correction tends to zero as $0.035
\: {e^2 / \hbar} \: \sqrt{B_{tr} / B}$.

\subsection{Interplay of Rashba and Dresselhaus terms}

Here we study the effect of interference of Dresselhaus and Rashba spin-orbit interactions on weak
localization. We assume that, in the effective field Eq.~\eqref{Omega}, the
first angular harmonics ${\bm \Omega}{^{(1)}}({\bm k})$ prevail over ${\bm \Omega}{^{(3)}}({\bm k})$.

In the presence of both $\bm k$-linear Dresselhaus and Rashba spin-splittings the system has C$_{\rm 2v}$ point symmetry which does not allow even {a} block-diagonalization of the operators $\mathcal A$ and $\mathcal K$. In this case the calculations are performed {numerically} by using Eqs.~\eqref{sigma_a}-\eqref{A_NN1}, see Ref.~\cite{FTP06} for details.

\begin{figure}
\includegraphics[width=\linewidth]{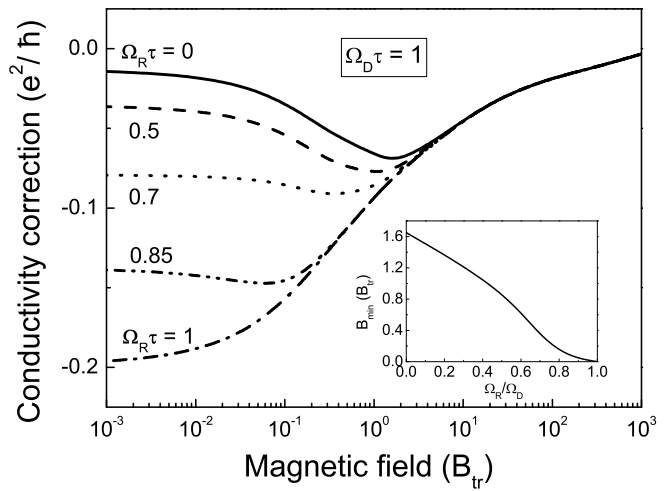}
\caption{Quantum corrections to the conductivity calculated  at the fixed magnitude
of the first harmonic of Dresselhaus component, $\Omega_D\tau = 1$, for different values of Rashba constant $\Omega_R\tau$ ($\tau/\tau_\phi = 0.01$). The inset shows the
dependence of the position of the minimum in the curve for
the magnetoconductivity on the ratio $\Omega_R/\Omega_D$. After~\cite{FTP06}.}
 \label{Interplay_RD}
\end{figure}

Figure~\ref{Interplay_RD} shows the dependence of the weak localization
correction to the conductivity on magnetic
field 
%for different ratios between the Dresselhaus term and Rashba term. 
assuming that the magnitude of the first
harmonics of the Dresselhaus term is constant, $\Omega_D\tau = 1$.
Different curves refer to different ratios between
Rashba and Dresselhaus contributions
($\Omega_R/\Omega_D = 0, 0.5, 0.7, 0.85$, and 1.0). 

We begin to analyze the results with the case of
$\Omega_R=\Omega_D$. In this limit, energy spectrum splits into two independent paraboloids. Each of them provides {the} universal contribution to the magnetoresistance, so the total correction is the same as in the absence of spin-orbit interaction, it is  
described by Eqs.~\eqref{sigma_a_no_spin},~\eqref{sigma_b_no_spin}. {The effective field $\bm \Omega(\bm k)$ in this situation points along the fixed axis, therefore spin rotation angle on the closed loop is zero.}
The absolute value of the quantum correction to the conductivity
steadily increases. In high fields ($B/B_{tr} \gg 1$), the
correction is described by the asymptotic expression~\eqref{hf_asympt}.

With unequal Rashba and Dresselhaus terms, the rotation angle of the spin {at} closed paths is no
longer zero. 
This yields (i) a smaller magnitude
of the correction and (ii) an alternative
magnetoresistance. Since, for a given path, the rotation
angle of the spin is larger the larger is the quantity
$|\Omega_R^2 -
\Omega_D^2|\tau^2$, a decrease in Rashba term (at a fixed
Dresselhaus term) manifests itself as an enhancement
of the spin-orbit coupling. In fact, as is evident from
Fig.~\ref{Interplay_RD}, ``the depth'' of the minimum increases with
decreasing ratio $\Omega_R/\Omega_D$, and the minimum itself shifts
to higher magnetic fields. These results are in qualitative
agreement with the diffusion theory~\cite{PikusPikus}. In contrast
to Ref.~\cite{PikusPikus}, the theory developed here provides a correct
asymptotic behavior of the quantum correction to the
conductivity at $B\gg B_{tr}$: irrespective to the quantity
$\Omega_R/\Omega_D$, all curves approach the same dependence~\eqref{hf_asympt}.

\subsection{Cubic in momentum splitting}

\begin{figure}
\includegraphics[width=\linewidth]{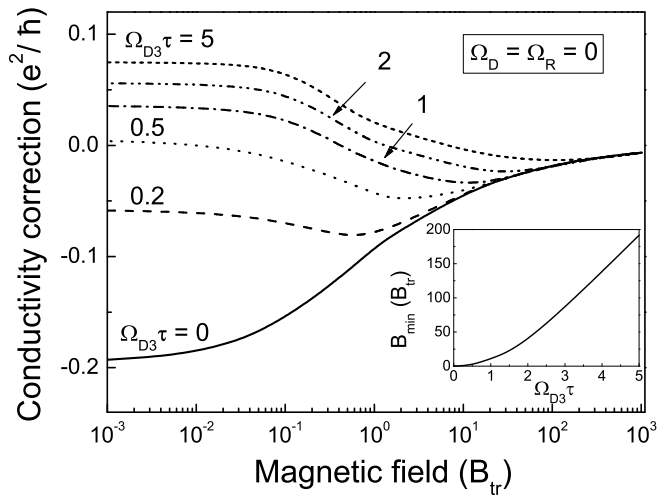}
\caption{Dependences of the conductivity on magnetic
field for different magnitudes of the third
harmonic of Dresselhaus component, $\Omega_{D3}\tau$, at $\tau/\tau_\phi=0.01$. The inset shows the position of the minimum as a function of $\Omega_{D3}\tau$. After~\cite{FTP06}.}
 \label{k_cubic}
\end{figure}

Here we discuss the situation where the third harmonic ${\bm \Omega^{(3)}_D({\bm k})}$ is the only one in
the energy spectrum. This situation is relevant to $p$-type heterostructures and, to a large extent, to the case of the equal first harmonics of Dresselhaus contribution and
Rashba contribution {($\Omega_R = \Omega_D$)}.

If ${\bm \Omega^{(3)}_D({\bm k})}$ is dominant then the spin splitting is isotropic in $\bm k$-space. Therefore, the operators $\mathcal A$ and $\mathcal K$ are again separated into blocks of sizes $3\times 3$.
The third harmonic of Dresselhaus
contribution mixes the states with equal values
of $N+3m$. The corresponding expressions for the quantum correction to the conductivity are given in Ref.~\cite{FTP06}.

Figure~\ref{k_cubic} shows the dependences of the conductivity correction 
on the magnetic field, as calculated for different
magnitudes of the third harmonic of Dresselhaus
term, with the first harmonic $\bm \Omega^{(1)}{(\bm k)}$ equal to zero. Qualitatively, the form of the dependences is consistent
with the results for the spin splitting described by
the first harmonic, Fig.~\ref{IsotrSS}, but the range of variation of $\sigma(B)$ is about an order of magnitude larger. In the magnetic fields $B\gg
\max{[(\Omega_{D3}\tau)^2,1]}B_{tr}$, all curves approach the same zero-$\Omega$
dependence~\eqref{hf_asympt}. With increasing $\Omega_{D3}\tau$, the minimum
of the conductivity shifts to higher fields. The
depth of the minimum behaves nonmonotonically: with
increasing spin splitting, it increases at small $\Omega_{D3}\tau$ and decreases at large $\Omega_{D3}\tau$.

\subsection{Comparison with experiment}

\begin{figure*}
 \includegraphics[width=0.48\textwidth]{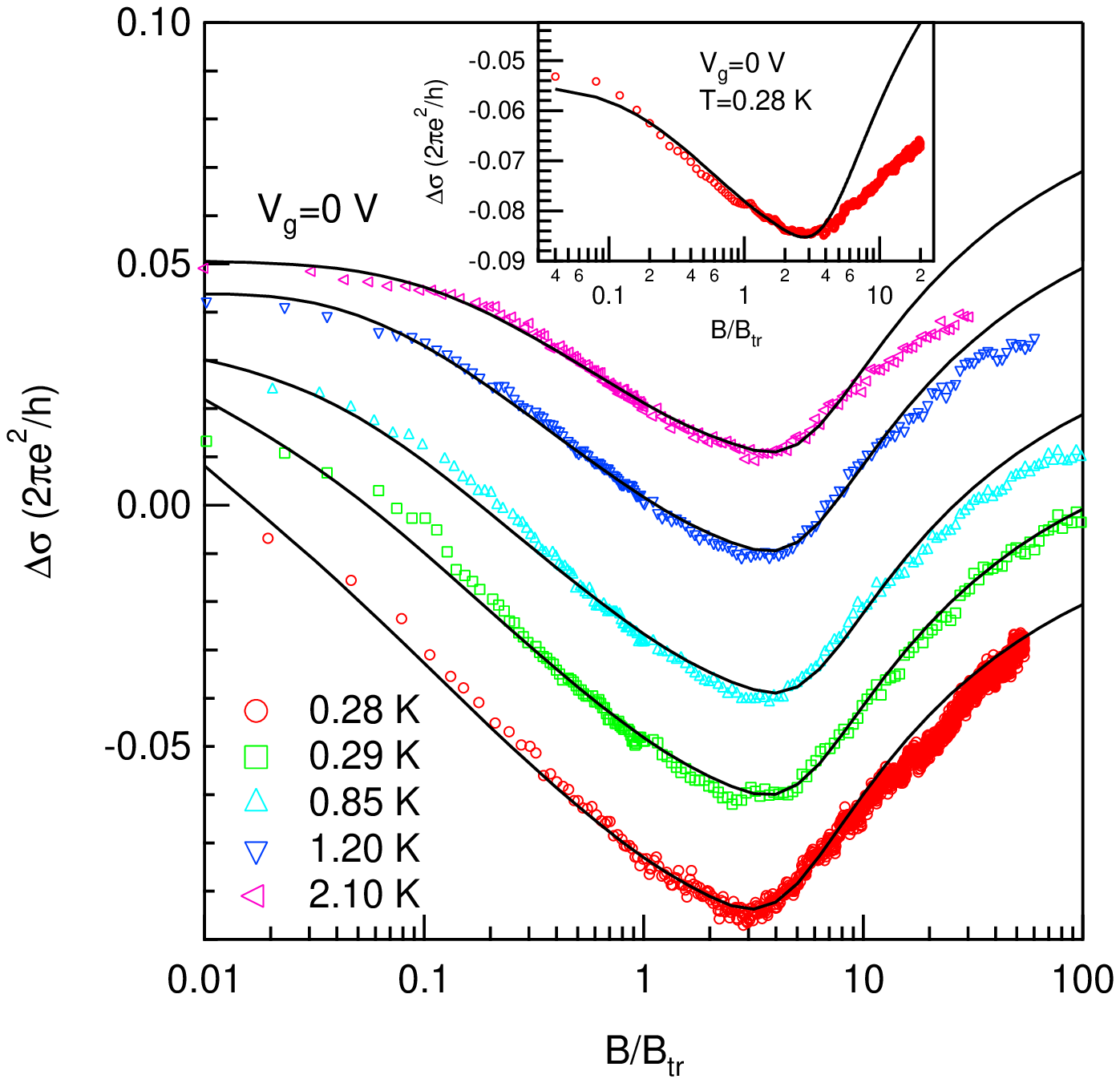}
 \includegraphics[width=0.48\textwidth]{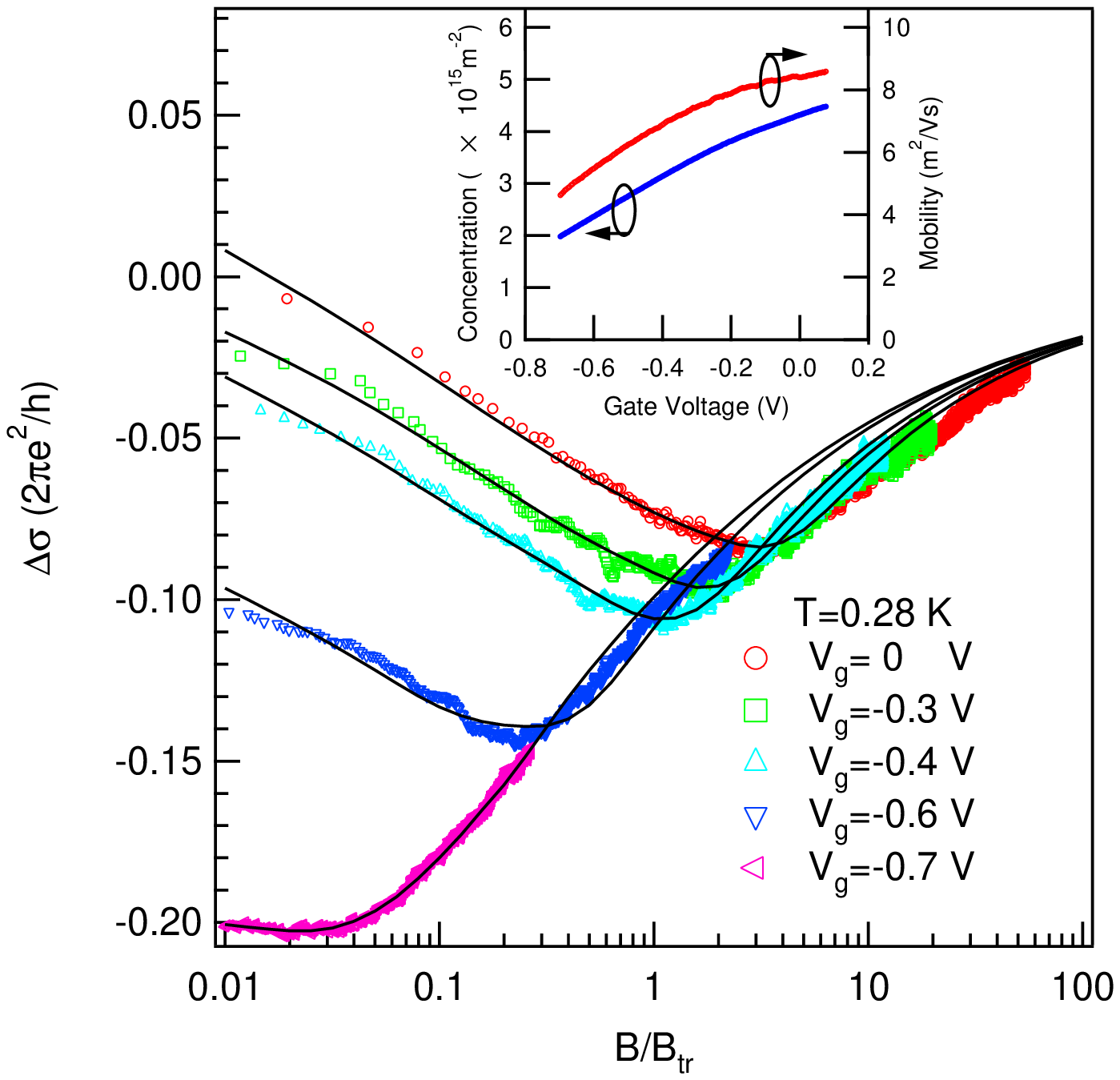}
\caption{Conductivity correction versus magnetic field. Points are experimental data, solid curves are 
theoretical fits. Left panel: magnetic field dependences for 
different temperatures at a fixed gate voltage $V_g=0$.  Inset shows the fit by diffusion theory which is unsatisfactory in high magnetic fields. Right panel: conductivity dependences on magnetic 
field for different gate voltages at a fixed temperature $T = 0.28$~K. 
Inset presents measured gate voltage dependences of carrier 
concentration and mobility. After~\cite{exp_posle7}.
}
 \label{Stud}
\end{figure*}

The theory developed in Refs.~\cite{PRB05,FTP06} has been successfully applied for description of experimental data on anomalous magnetoresistance in various high-mobility heterostructures. For the first time it has been done in the work by \textit{Guzenko~et~al.}~\cite{exp_posle1} where the parameter $\Omega\tau$ as well as the temperature dependence of the dephasing time $\tau_\phi$ have been determined {from the} fitting of the magnetoconductivity curves. Note that the {experimental }data {were} fitted in a temperature range 0.8\ldots4~K by the same parameter $\Omega\tau \approx 0.5$ for magnetic fields $B \leq 8\: B_{tr}$ while the minimum took place at $B \approx 2\: B_{tr}$. The extracted  spin-orbit splitting $\approx 1$~meV is a reasonable value for the Rashba splitting for the studied  2D electrons in InGaSb heterostructures. Note that this splitting could not be determined e.g. by beatings in the Shubnikov-de~Haas oscillations because it has an order of $\hbar/\tau$. At the same time the anomalous magnetoconductivity had a pronounced minimum allowed extraction of the Rashba spin splitting.

Later on the theory has been applied independently in different experimental groups allowing characterization of electron spin properties in various 2D semiconductor heterostructures, see e.g.~\cite{exp_posle3,exp_posle4,exp_posle5,exp_posle6,exp_posle7}. 
In the work by \textit{Yu~et~al.}~\cite{exp_posle7} the structures with an electric gate were investigated with allowance for a change of concentration and mobility of 2D electrons.
The results of this study are summarized in Fig.~\ref{Stud}. Left panel  
of Fig.~\ref{Stud} shows the conductivity correction as a function of 
magnetic field plotted for different temperatures at a fixed gate 
voltage. All curves have minima at $B_{min}\sim 2B_{tr}$ and are well 
reproduced by the theoretical fit according to Eqs.~\eqref{sigma_a_fin}, 
\eqref{sigma_b_fin}. An inset to the left panel shows low-temperature data 
fitted by a diffusion theory, see Ref.~\cite{exp_posle7} for details. It 
is clearly seen that the diffusion theory does not reproduce high-field 
part of the magnetoconductivity.
Right panel of Fig.~\ref{Stud} shows conductivity correction as a function 
of magnetic field measured at different values of a gate voltage. In 
accordance with the inset to the right panel the variation of the gate 
voltage induces variation of the carrier concentration and mobility, 
therefore the spin splitting parameters $\Omega\tau$ is changed along 
with $\tau/\tau_{\phi}$. Figure~\ref{Stud} demonstrates that at high 
magnetic fields all curves reach the same universal asymptotics. 
Figure~\ref{Stud} evidences that the theory describes the experimental 
data up to the fields 50~$B_{tr}$ in the whole range of used 
temperatures and gate voltages.

\section{Effect of an in-plane magnetic field}

Magnetic field applied in the plane of the quantum well affects an interference of electrons in a two-fold way. First, inevitable fluctuations of the quantum well width caused by the imperfections of interfaces lead to the fluctuations of the magnetic flux through the cross-section of electron wave function. This effect leads to the dephasing similarly to the case of perpendicular field~\cite{Malsh1}. Further on this orbital or ``microroughness'' effect is neglected.

The other possibility for an in-plane magnetic field to affect weak localization or antilocalization of electrons is the Zeeman effect. The corresponding term in the electrons Hamiltonian is 
\begin{equation}
 \label{Zeeman}
H_{\rm Z} = \frac{\hbar}{2} \bm \sigma \cdot \bm \Delta,
\end{equation}
where $\bm \Delta = \bm {\hat g}\mu_B {\bm B}_\parallel/\hbar$, ${\bm B}_\parallel$ is the in-plane magnetic field, $\mu_B$ is the Bohr magneton, and $\bm {\hat g}$ is the in-plane electron 
Land\'{e}-factor tensor. For simplicity we consider here the situation where only isotropic spin splitting is present ($\Omega_R\ne 0$, $\Omega_D =0$).

 An in-plane magnetic field admixes triplet states of interfering electron pair to the singlet one, thus causing {a} dephasing of the singlet state~\cite{Malsh1,marinescu}. Theory~\cite{Malsh1} describes anomalous magnetoresistance in the presence of in-plane field~\cite{tilted}. However the expressions obtained in Refs.~\cite{Malsh1,marinescu} cover only the case of small spin splittings and relatively weak magnetic fields: $\Omega\tau\ll 1$, $\Delta \, (\tau \tau_\phi)^{1/2} \ll 1$. Below we present a general theory describing an in-plane magnetoresistance for the arbitrary values of the spin-orbit and Zeeman splittings.

The conductivity corrections in this case are given by
\begin{equation}\label{sigma_a_Bpar}
    \sigma_a (B_\parallel) = - {e^2 \over 2 \pi \hbar} l^2 
    \int {d^2q\over (2\pi)^2}
    \left[ \mathcal P^3 ({\cal I} - \mathcal P)^{-1} \right]_{\alpha\beta\beta\alpha},
\end{equation}

\begin{equation}\label{sigma_b_Bpar}
 \sigma_b(B_\parallel) = {e^2 \over \pi \hbar} l^2 
    \int {d^2q\over (2\pi)^2}
     \mathcal Q_{\delta\beta\mu\nu} \mathcal Q_{\nu\mu\gamma\alpha} \left[\mathcal P ({\cal I} - \mathcal P)^{-1}\right]_{\alpha\gamma\beta\delta}.
\end{equation}
Here 
\begin{eqnarray}
 \label{Aq}
\mathcal P ({\bm q}) = \int &d^2 R&  {\exp{(-R/\tilde{l})} \over 2\pi R l}  
\\
&\times& \exp{[{\rm i} {\bm q}\cdot{\bm R} - 2\mathrm i \bm S\cdot \bm \omega(\bm R) - 2\mathrm i \bm L\cdot \bm \Delta]},
\nonumber
\end{eqnarray}
and the operator $\mathcal Q ({\bm q})$ differs by an additional factor ${\rm i} X/R$ in the integrand~\eqref{Aq}.
$\bm L$ is the operator of spin difference of interfering particles. {Greek subscripts $\alpha\ldots\nu = \pm 1/2$ enumerate spin states of the interfering particles, the summation over repeated subscripts is assumed.} It is worth noting, that the Zeeman effect in interference is determined by difference of spins since Zeeman splitting is an even function of the wavevector, similarly to the effect of longitudinal-transverse splitting of exciton-polaritons~\cite{Polarit_PRB08}. Therefore singlet and triplet contributions are mixed by the in-plane magnetic field. 
This is a manifestation of C$_{\rm s}$ point symmetry of the system in the presence of both ${\bm B}_\parallel$ and Rashba spin-orbit interaction.

In contrast, for Zeeman splitting due to the field $B_z$ applied perpendicularly to the
2D plane the conductivity corrections have the simplified
form similar to Eqs.~\eqref{sigma_a_fin},~\eqref{sigma_b_fin}. Singlet and triplet states do not mix in this case because a presence of the $B_z$ component does not break an in-plane isotropy of the
energy spectrum. Formally this can be seen from Eq.~\eqref{Aq} noting
that the operator $L_z$ commutes with the operator $\bm S$.

\begin{figure}
\includegraphics[width=\linewidth]{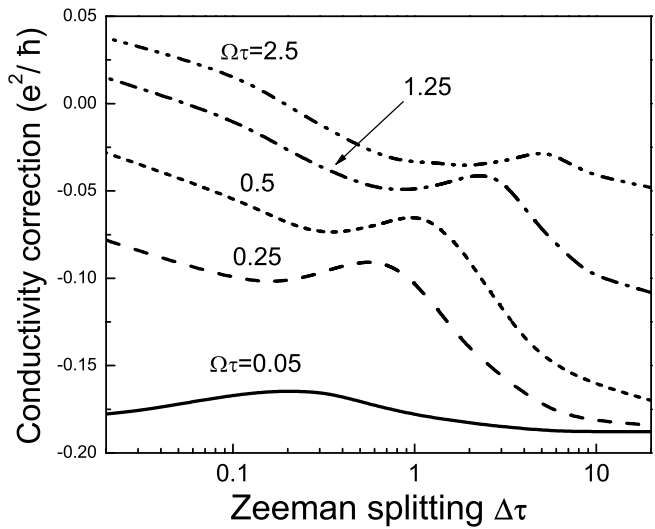}
\caption{Dependences of the conductivity correction on the in-plane magnetic
field due to the Zeeman effect for different magnitudes of the isotropic spin splitting, $\Omega\tau$, at $\tau/\tau_\phi=0.01$.}
 \label{Par_field}
\end{figure}

Before presenting numerical results let us briefly discuss the situation of small spin-orbit splittings of conduction band and high Zeeman splitting: $\Omega\tau\ll 1$, $\Delta\tau\gg 1$. In the absence of the spin-orbit interaction, Zeeman effect splits electron energy spectrum into two independent paraboloids. Provided that $\Delta/E_{\rm F}\ll 1$ the difference of their occupations and relaxation times can be neglected, and each paraboloid yields the same universal contribution to the quantum conductivity correction. As a result the latter equals to the spin-orbit-less value given by 
\begin{equation}
\label{spinless}
	\sigma(0)=- {e^2 \over 2 \pi^2 \hbar} \ln{\left({\tau_\phi\over2\tau}\right)}.
\end{equation}
Small spin-orbit interaction leads to the spin relaxation and mixing of paraboloids. However, at $\Delta\tau\gg 1$ and $\Delta\gg \Omega$ Zeeman effect of the magnetic field quenches spin relaxation completely~\cite{larmor}, therefore even in the presence of the spin-orbit interaction the in-plane magnetoconductivity approaches at high magnetic field $B_\parallel$ the spin-orbit-less value Eq.~\eqref{spinless}.

The quantum corrections to the conductivity calculated by Eqs.~\eqref{sigma_a_Bpar} and~~\eqref{sigma_b_Bpar} are shown in Fig.~\ref{Par_field}. Different curves correspond to the different values of spin-orbit splitting. The dependences start from the zero-field values of conductivity correction at a corresponding spin splitting~\cite{PRB05} and, at $\Delta\tau \gg 1$, tend to the spin-orbit-less value Eq.~\eqref{spinless}. In small in-plane fields, $\Delta\tau \ll 1$, the conductivity correction decreases because the singlet contribution is suppressed due to an admixture of triplet states. This situation is described by the theory developed in Refs.~\cite{Malsh1,marinescu}. However, in the intermediate area $\Delta\sim \Omega$ the spectrum of interfering states is rearranged. It results in non-monotonous dependence of the conductivity correction on the magnetic field. First, Zeeman splitting supresses the singlet thus decreasing the conductivity. With an increase of $\Delta$, triplet states start to be supressed as well and the conductivity increases. At very high fields the Zeeman splitting completely overcomes the spin-orbit splitting and restores the spin-less situation thus decreasing the conductivity again.
{This results in the presence of minimum and maximum in the dependence $\sigma(B_\parallel)$ at $\Delta \sim \Omega$, Fig.~\ref{Par_field}, absent in the theories~\cite{Malsh1,marinescu}. It is noteworthy that contrary to Refs.~\cite{Malsh1,tilted} the non-monotonous behaviour of the conductivity correction takes place in the absence of the ``microroughness'' effect, i.e. due to interference of 
spin effects only.}

Recently the diffusion theory has been applied for calculation of a low-field magnetoconductivity of thin quantum wires~\cite{richnit}. The theory developed here allows one to describe adequately the weak localization effect even in ballistic structures with dimensions comparable with the mean free path.

\section{Conclusion}

To summarize, {we have presented a theory of quantum conductivity corrections valid for the state-of-the-art semiconductor heterostructures. The theory is applicable in the whole range of classically weak magnetic fields allowing for both the diffusive and ballistic carrier propagation. It takes into account all possible contributions to the spin splitting of electron energy spectrum such as linear in the wavevector Rashba and Dresselhaus terms and cubic in the wavevector Dresselhaus term. We have demonstrated that in the case of equal Rashba and Dresselhaus terms the magnetoconductivity shows monotonic behaviour contrary to the general case of unequal Rashba and Dresselhaus contributions or dominant cubic-in wavevector splitting. The theory is shown to be in the excellent agreement with the recent experimental data.}

{We have also studied the magnetoconductivity in the in-plane field. Although its orbital effects are relatively unimportant, the Zeeman effect of the field enters in the competition with {the} conduction band spin-orbit splitting. We predict in high-mobility heterostructures an alternating magnetoconductivity due to the interplay of various spin splittings.}

\section*{Acknowledgements}

This work was financially supported by the RFBR, 
%\addMisha{Programmes of RAS}, 
``Dynasty'' Foundation --- ICFPM, and President grant for young russian scientists.

\par
\end{document}